\documentclass{INTERSPEECH2023}
\usepackage{amsmath, amssymb}
\usepackage{times}
\usepackage{soul}
\usepackage{url}
\usepackage{booktabs}
\usepackage{bm}
\usepackage{bigstrut}
\usepackage{amsmath}
\usepackage{amsfonts}       
\usepackage{nicefrac}       
\usepackage{microtype}      
\usepackage{xcolor}         
\usepackage{multirow}
\usepackage{graphicx}
\usepackage{subfigure}
\usepackage{hyperref}
\usepackage{wrapfig}

\usepackage{algorithmic}
\usepackage[ruled]{algorithm2e}

\SetKwInput{KwInput}{Input}
\SetKwInput{KwOutput}{Output}

\usepackage{indentfirst}

\usepackage{booktabs}  
\usepackage{threeparttable}  
\usepackage{multicol}  
\usepackage{multirow}  
\usepackage{xspace}
\usepackage{colortbl}

\def\eg{\emph{e.g.}}
\def\ie{\emph{i.e.}}

\newcommand{\wapat}{\textsc{Wapat}\xspace}

\interspeechcameraready

\title{Robust Automatic Speech Recognition via WavAugment Guided Phoneme Adversarial Training}
\name{Gege Qi$^1$, Yuefeng Chen$^1$, Xiaofeng Mao$^1$, Xiaojun Jia$^2$, Ranjie Duan$^1$, Rong Zhang$^1$, Hui Xue$^1$}
\address{
  $^1$Alibaba Group, China\\
  $^2$Chinese Academy of Sciences, China}
\email{\{qigege.qgg,yuefeng.chenyf,mxf164419\}@alibaba-inc.com}
\begin{document}

\maketitle

\begin{abstract}
Developing a practically-robust automatic speech recognition (ASR) is challenging since the model should not only maintain the original performance on clean samples, but also achieve consistent efficacy under small volume perturbations and large domain shifts. To address this problem, we propose a novel WavAugment Guided Phoneme Adversarial Training (\wapat).
\wapat use adversarial examples in phoneme space as augmentation to make the model invariant to minor fluctuations in phoneme representation and preserve the performance on clean samples.
In addition, \wapat utilizes the phoneme representation of augmented samples to guide the generation of adversaries, which helps to find more stable and diverse gradient-directions, resulting in improved generalization.
Extensive experiments demonstrate the effectiveness of \wapat on End-to-end Speech Challenge Benchmark (ESB).
Notably, SpeechLM-\wapat outperforms the original model by 6.28\% WER reduction on ESB, achieving the new state-of-the-art.


\end{abstract}

\section{Introduction}

Nowadays, there have been remarkable advancements in Deep Neural Network (DNN) based Automatic Speech Recognition (ASR)~\cite{baevski2020wav2vec}, resulting in the emergence of numerous speech-related applications that assist humans in their daily activities. 
However, despite the impressive performance of ASR systems, they are limited to specific tasks since they assume that the training and testing data are drawn from the same distribution~\cite{parada2022pmct}.
Thus, applying ASR in real-world applications under diverse environment is still a huge challenge ~\cite{hu2022dual,fan2022draft,hu2023gradient}.

In this work, we aim to address such a challenging cross-domain scenario where an ASR system needs to be robust against various potential distortion. However, there are two major challenges:
  1) \textbf{Robustness against perturbation:} Real-world volume perturbation (\eg, environmental noise, reverberation, and background speakers) significantly impacts the performance of an ASR system~\cite{gajic2006robust,ko2017study}.
  2) \textbf{Robustness Generalization:} There exist various type of volume perturbation in practical scenario. However, a ASR is robust against one type of perturbation not promised being robust under unknown domain (\eg, change of speaking style). 
Existing work either adopt data augmentation to improve ASR's robustness against specific perturbation but limited under unseen domain \cite{jaitly2013vocal, peddinti2015time, park2019specaugment, ko2015audio}, or use speech enhancement as a pre-processing to deal with various potential noise \cite{tan2020improving}. Both of them fail to achieve a real-robust ASR system which can be applied in the real world. Therefore, enhancing the robustness of ASRs while improving their robustness generalization across different perturbation remains a significant challenge.

\begin{figure*}[t]
  \centering
  \includegraphics[width=\linewidth]{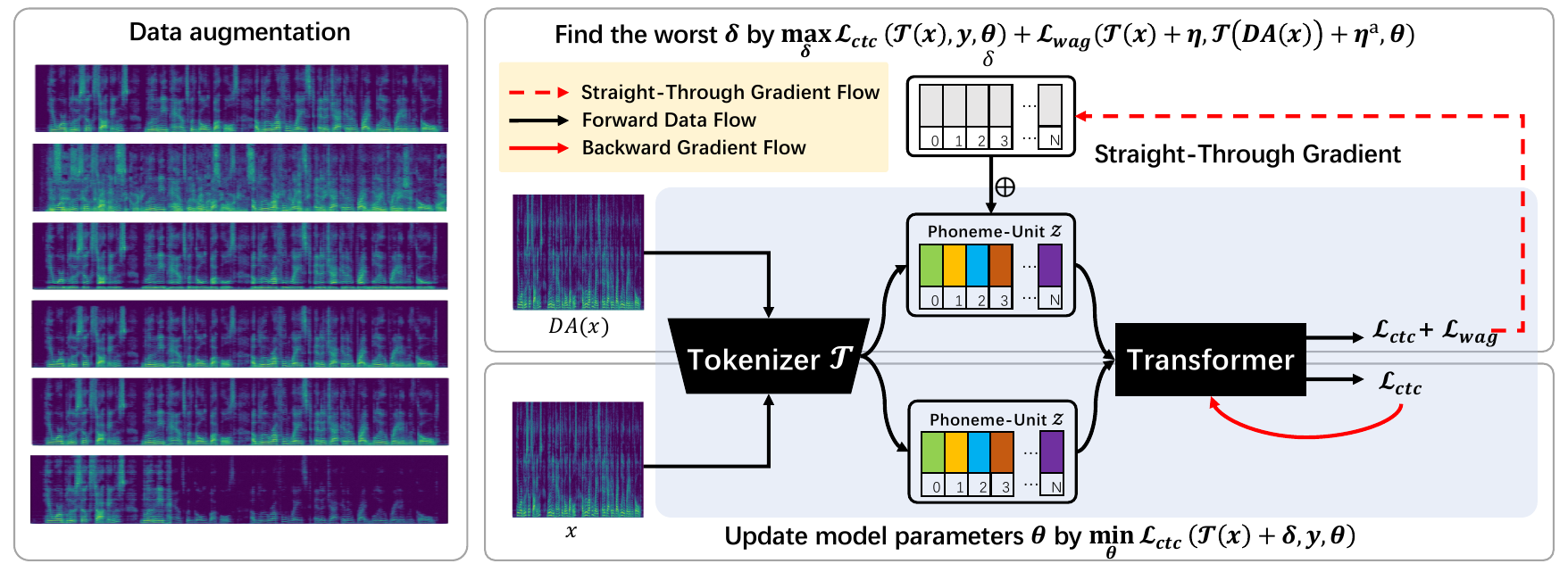}
  \caption{The overview of our proposed WavAugment Guided Phoneme Adversarial Training (\wapat). Left: from top to bottom, the figures depict the log mel spectrogram of the base input with no augmentation, additive noise, band reject, pitch modification, time masking and reverberation applied. Right: the pipeline of \wapat, where the augmented samples are used to guide the generation of adversaries during adversarial training on the symbolic phoneme space.
  }
  \label{fig:3_1}
  \vspace{-0.4cm}
\end{figure*}

To address the challenge, we propose a novel method called \textbf{W}av\textbf{A}ugment Guided \textbf{P}honeme \textbf{A}dversarial \textbf{T}raining (\wapat) by leveraging adversarial training (AT) technique. Though previous work \cite{zhang2019theoretically} claimed AT results in a trade-off between robustness and clean accuracy. However, research in the field of natural language processing~\cite{ivgi2021achieving} and recent in computer vision~\cite{mao2022enhance} demonstrated that aligning the distributions of adversarial and original samples during AT can benefit robustness and clean accuracy simultaneously. Borrowing the idea, we propose applying AT on phoneme space to create adversarial speech with realistic semantic. To be specific, \wapat employs a single-step attack to generate adversarial perturbations on phoneme representations. A data augmentation is further applied to guide the attack for generating more stable and diverse phoneme adversarial examples.
In detail, with the time-domain WavAugment~\cite{kharitonov2021data} technique, we use Kullback-Leibler Divergence (KLD) to align the adversaries on original samples and those on augmented samples. Furthermore, multiple augmentations in WavAugment are used to guide the adversarial training for searching for diverse gradient directions and leading to better generalization.
Figure~\ref{fig:3_1} shows the overall pipeline of WavAugment guided phoneme adversarial training.

We further explore the impact of individual techniques in WavAugment and their combinations with \wapat on the performance of the model.
Our findings indicate that while hard augmentations can improve robustness on some datasets, they fails on others.
It is reasonable to expect challenges in generalizing audio augmentations across different domains, given the inherent complexity of audio signals.
Instead, our \wapat consistently improves performance in terms of both cross-domain datasets and different types of transformations.
The stability of generalization indicates that the WavAugment-guided adversary is effective in inducing robust features into target ASR.

In summary, we make the following contributions:
\begin{itemize}
    \item To our knowledge, this is the first work that sheds light on adversarial training on phoneme-unit space for improving standard performance and generalization simultaneously.
    \item We propose WavAugment Guided Phoneme Adversarial Training (\wapat), which employs phoneme representation of the augmented audios to guide the generation of adversaries, resulting in more diverse robust features.
    \item By combining SpeechLM~\cite{zhang2022speechlm} pre-training and \wapat fine-tuning, our method achieves new state-of-the-art performance on the challenging benchmark ESB~\cite{gandhi2022esb}, which contains multiple speech datasets from a broad range of domains.
\end{itemize}

\section{Related Work}
Existing approaches for improving ASR generally from two aspects: \textbf{1) Improving ASR's robustness against specific perturbation:} Early works have shown that several data augmentation methods, such as vocal tract length perturbation~\cite{jaitly2013vocal}, volume perturbation~\cite{peddinti2015time} and speed perturbation~\cite{ko2015audio}, can improve the robustness of ASR models.
SpecAugment~\cite{park2019specaugment} is widely used to train ASR models due to its efficiency.
Specifically, SpecAugment randomly masks chunks of time or frequency channels on spectrograms.
However, these DA techniques are typically designed manually for specific domains based on domain-specific knowledge and experience.
When dealing with an unknown target domain or multiple domains, it can be challenging for experts to apply specific transformations, or to construct and fine-tune more sophisticated augmentation compositions~\cite{damania2022combining}.
Besides data augmentation, several work utilize adversarial training \cite{sun2018training, yang2020characterizing} aiming to improve ASR's adversarial robustness under adversarial examples. However, all of these work target ASR's robustness under specific perturbation, and the improved ASR is still limited on unseen domain.
\textbf{2) Improving ASR's performance via pre-processing:} To achive better performance, there are also several works propose using speech enhancement methods to remove noise from speech signals before passing them through a standard ASR~\cite{tan2020improving}. A more recent approach that operates on raw waveform for real-time speech enhancement is~\cite{defossez2020real}. However, these methods often rely on front-end processing modules, which decrease efficiency and add computational overhead. Also, the speech enhancement method do not really improve the robustness of ASR itself.
In this paper, we aim to build a truly robust ASR which is robust under multiple or even unseen perturbations. It has the potential to be applied in various applications in real-world setting.

\section{Method}
\subsection{Adversarial Training on ASR}
Consider the training utterance and text label set $\mathcal{D}=\{(x_i, y_i)_{i=1}^n\}$, an ASR model with learnable parameters $\theta$, and a recognition objective given by Connectionist Temporal Classification (CTC) loss $\mathcal{L}_{ctc}$. Adversarial Training (AT) aims to optimize $\theta$ by solving a minimax optimization problem:
\begin{align}
\min_{\theta} \mathbb{E}_{(x, y) \sim \mathcal{D}} \left[ \max_{\delta} \mathcal{L}_{ctc}(x+\delta, y, \theta) \right] \ s.t. ||\delta||_p \leq \epsilon,
\label{equation:eq1}
\end{align}
where the inner optimization seeks perturbations $\delta$ on speech values that maximize the loss, and the outer minimization update $\theta$ to improve the worst-case performance of the network. The boundary $||\delta||_p \leq \epsilon$ restricts the magnitude of the perturbation.
We use projected gradient descent (PGD)~\cite{madry2017towards} for the inner optimization.
In the following section, we will tackle the challenges of mitigating performance degradation and enhancing the generalization ability of ASR models through AT techniques. 





\subsection{Phoneme Adversarial Training}

We borrow the perspective of the AT on the contextualized language representation, and propose a new Phoneme Adversarial Training (PAT) for ASRs, \ie, conducting AT on the phoneme representation space instead of raw input space.
To accomplish this, we leverage the SpeechLM framework proposed by~\cite{zhang2022speechlm} to recognize speeches. The phoneme unit sequence of input $x$ can be obtained by applying a transformer based phoneme-unit tokenizer $\mathcal{T}$.
In the inner maximization step of AT, we generate phoneme adversarial examples by slightly modifying Equation~\ref{equation:eq1}. The objective of PAT can be formulated as follows:
\begin{align}
\min_{\theta} \mathbb{E}_{(x, y) \sim \mathcal{D}} \left[ \max_{\delta} \mathcal{L}_{ctc}(\mathcal{T}(x)+\delta, y, \theta) \right] \ s.t. ||\delta||_{\infty} \leq \epsilon,
\label{equation:eq3}
\end{align}
where $\epsilon$ is the magnitude of the perturbation, PAT finds the failure case, \ie, $\mathcal{T}(x)+\delta$ in phoneme space.

\IncMargin{-1.5em} 
\begin{algorithm}[ht]
\caption{Pseudo code of \wapat}
\label{alg:awat}
\KwInput{Speech tokenizer $\mathcal{T}$; A sampled mini-batch of clean audios $x$ with labels $y$; Perturbation size $\epsilon$.}
\KwOutput{Learned network parameter $\theta$} 
\begin{flushleft}
\begin{algorithmic}[1]
\raggedright
\STATE Fix the network parameters of $\mathcal{T}$
\FOR{each training step}
\STATE $z \leftarrow \mathcal{T}(x)$
\STATE $z \leftarrow \mathcal{U}(\mathcal{B}_{\epsilon}^{\infty}(z))$\hfill\COMMENT{Initialize adversarial example}
\STATE $z^{a} \leftarrow \mathcal{T}(\mathcal{DA}(x))$
\STATE $\eta, \eta^a \leftarrow \nabla_{z} \mathcal{L}_{ctc}(z,y,\theta), \nabla_{z} \mathcal{L}_{ctc}(\mathcal{T}(z^a,y,\theta)$
\STATE $\delta \leftarrow \nabla_{z} [\mathcal{L}_{ctc}(z,y,\theta) + \mathcal{L}_{wag}(z+\eta, z^{a}+\eta^a,\theta)]$
\STATE $\hat{z} \leftarrow \prod\limits_{\mathcal{B}_{\epsilon}^{\infty}(z)}(z+ \delta)$\hfill\COMMENT{Generate adversarial examples}
\STATE Update model parameter on $\mathcal{L}_{ctc}(\hat{z},y,\theta)$
\ENDFOR
\end{algorithmic}
\end{flushleft}
\end{algorithm}

\subsection{WavAugment Guided Phoneme Adversarial Training}
For improving the generalization of ASR by AT, we aim to generate adversarial examples that exhibit both stability and diversity.
Towards this end, we propose a novel WavAugment Guided Phoneme Adversarial Training (\wapat) method.

Adversarial examples are typically distributed near the decision boundary, and slight variations can cause them to lose their adversarial nature~\cite{goodfellow2014explaining}.
Enhancing the stability of adversarial examples is beneficial for obtaining more robust features.
To tackle the instability problem, we introduce the WavAugment guided term along with the CTC loss to form a new objective function during the generation of adversarial examples.

The WavAugment operation, denoted as $\mathcal{DA(\cdot)}$, applies time-domain data augmentation to an audio sample. 
We represent the phoneme representation of the original sample and the augmented sample as $z$ and $z^a$, respectively. 
Then, the perturbations generated for the two samples are denoted as $\eta$ and $\eta^a$.
The loss WavAugment guided term encourages the predictions of the adversarial examples of the original and augmented samples to be similar.
Formally, the objective function can be written as:
\vspace{-0.5em}
\setlength{\belowdisplayskip}{-0.000000000005em}
\begin{align}
\mathcal{L}_{wag}(z+\eta,z^a+\eta^a,\theta)&\!=\! -\mathcal{D}_{KL}\left[p(z+\eta,\theta)||
p(z^a+\eta^a,\theta)\right] \nonumber\\
z=&\mathcal{T}(x), z^a=\mathcal{T}(\mathcal{DA}(x))
\end{align}
where $p(x|\theta)$ is the joint probability and $\mathcal{D}_{KL}$ is the KL-divergence.
From an optimization perspective, the WavAugment guided term helps in avoiding local optima during the perturbation generation process, leading to the creation of more stable and robust features for ASR models.



In this paper, the basic data augmentations of WavAugment~\cite{kharitonov2021data} is reserved, including pitch modification (\texttt{pitch}), additive noise (\texttt{add}), band reject filtering (\texttt{band\_rej}), time masking (\texttt{time\_mask}) and reverberation (\texttt{reverb}).
\texttt{pitch} and \texttt{add} are intended to simulate variations in the speaker's voice and environmental noise.
\texttt{band\_rej} and \texttt{time\_mask} augmentations can introduce noise into the neural representation of speech, which can help the model learn to better handle noisy speech.
The \texttt{reverb} simulates the effect of sound reflections in a room, which can help the model learn to better handle the effects of reverberation in real-world environments.
Here, we use gpuRIR~\cite{diaz2021gpurir} to obtain acoustic room impulse responses.

To enhance the diversity of adversarial examples, we utilize all of the augmentations available in WavAugment to guide the generation process. 
During training, one of the transformations from WavAugment is applied to each batch of samples.
In Figure~\ref{fig:3_1}, we show an example of log mel spectrograms augmented with different transformations.
Further details regarding the \wapat can be found in Algorithm~\ref{alg:awat}.
Given a SpeechLM based speech recognition model, the speech transformer $\mathcal{T}$ first yields a higher level phoneme representation $z$ from speech input. The WavAugment guided perturbation $\delta$ can be obtained by computing the gradient of $z$ towards maximizing the $\mathcal{L}_{ctc}$ and $\mathcal{L}_{wag}$. For clarity, $\mathcal{B}_{\epsilon}^{\infty}(z):=\{z^{\prime}:||z^{\prime}-z||_{\infty} \leq \epsilon\}$ defines a ball of radius $\epsilon$ around $z$ in the $l_\infty$ norm. The symbol $\mathcal{U}$ denotes the uniform distribution, and $\prod$ denotes a projection function. 
Finally, the adversarial example $\hat{z}$ is fed into models for training.

\begin{table*}[t]
\centering
\caption{WER comparison on the ESB benchmark over various methods for enhancing the robustness of ASR. Best performances are highlighted in bold.}
\vspace{-0.1cm}
\resizebox{1\hsize}{!}{
\begin{tabular}{l|cc|c|c|c|c|c|c|c|c|cc}
\toprule
{\multirow{2}*{\textbf{Method}}} & \multicolumn{2}{c|}{\textbf{Librispeech}} & {\multirow{2}*{\textbf{Chime-4}}}  & {\multirow{2}*{\textbf{Common Voice}}}  & {\multirow{2}*{\textbf{VoxPopuli}}}&  {\multirow{2}*{\textbf{TED-LIUM}}} &  {\multirow{2}*{\textbf{GigaSpeech}}} & {\multirow{2}*{\textbf{SPGISpeech}}}  &  {\multirow{2}*{\textbf{Earnings-22}}} &  {\multirow{2}*{\textbf{AMI}}} & {\multirow{2}*{\textbf{ESB score}}} \\
     ~ & \textbf{test-clean} & \textbf{test-other} & ~ & ~ & ~ & ~ & ~ & ~ & ~ & ~\\
\midrule
SpecAugment~\cite{park2019specaugment}  & 3.32 & 7.34 &	45.49 &	38.46 &	36.47 &	19.03 &	24.57 &	20.10 &	51.10 &	45.02 &	36.19 \\
WavAugment~\cite{kharitonov2021data}  & 3.34 & 7.35 & 35.65 & 38.16 & 36.64 & 18.12 & 25.53 & 18.99 & 52.79 & 46.10 & 34.18 \\
AdvEx~\cite{sun2018training}     & 3.36 & 7.36 & 46.10 &	38.35 & 36.74 &	18.18 &	24.49 &	19.36 &	52.01 &	44.79 &	36.24  \\
DEMUCS~\cite{defossez2020real}      & 3.33 & 7.29 &	33.57 &	43.63 &	36.71 & 18.31 & 24.39 &	26.24 &	56.76 &	44.63 &	35.32\\
\wapat        & \textbf{3.32} & \textbf{7.28} &	\textbf{32.68} &	\textbf{36.43} &	\textbf{36.38} &	\textbf{18.12} & \textbf{24.25} & \textbf{18.40} &	\textbf{49.78} & \textbf{44.53} &	\textbf{32.58} \\
\bottomrule
\end{tabular}}
\label{tab:overall}
\vspace{-0.2cm}
\end{table*}

\begin{figure*}[t]
  \centering
\includegraphics[width=\linewidth]{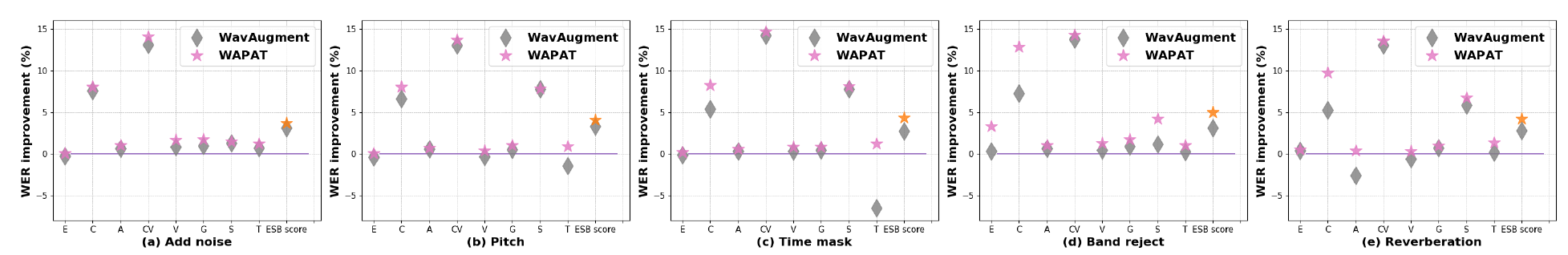}
 \vspace{-0.5cm}
  \caption{Comparison of the WER improvement on ESB benchmark, including Earnings-22 (E), CHiME-4 (C), AMI (A), Common Voice (CV), VoxPopuli (V), GigaSpeech(G), SPGISpeech (S) and TED-LIUM (T) dataset. The last column is the ESB score.
  }
  \label{fig:3_2}
  \vspace{-0.4cm}
\end{figure*}

\section{Experiment}
\label{sec:exp}
\noindent \textbf{Datasets and Settings}
We conducted experiments on the ESB~\cite{gandhi2022esb} benchmark to evaluate cross-domain ASR robustness. 
ESB comprises eight datasets with a broad range of domains, acoustic conditions, speaker styles, and transcription requirements. Notably, Librispeech~\cite{panayotov2015librispeech} and Common Voice~\cite{ardila2019common} only contain narrated style speeches, while VoxPopuli~\cite{wang2021voxpopuli} and TED-LIUM~\cite{rousseau2012ted} have oratory style speeches, and AMI ~\cite{carletta2007unleashing} contains spontaneous style speeches. GigaSpeech~\cite{chen2021gigaspeech}, SPGISpeech~\cite{o2021spgispeech}, and Earnings-22~\cite{del2022earnings} cover two different styles of speeches. Additionally, We included the optional CHiME-4~\cite{vincent2017analysis} dataset with narrated style to test generalization. We use the standard split of the above datasets and unify the transcription format as normalised.
We finetune the SpeechLM-P model\footnote{\url{https://github.com/microsoft/SpeechT5/tree/main/SpeechLM}} on the Librispeech-100h dataset, which is pre-trained on both the LibriSpeech-960h audio and the LibriSpeechLM corpus2. And we evaluate the robustness on datasets in ESB. 
Audio format is 16-bit WAV with 16 kHz, and transcription format is unified into the normalized form.

\noindent \textbf{Implementation Details}
The hyper-parameters used in WavAugment are as follows: 
\texttt{pitch} randomly modifies the pitch of the waveform by $n\in[-300,300]$ semitones.
\texttt{add} randomly adds noise from MUSAN~\cite{snyder2015musan} dataset with a scaled signal-to-noise ratio between $[0,40]$.
The maximal width of the rejected spectrum in \texttt{band\_rej} is 150 Hz. 
The \texttt{time\_mask} operation zeros out ten random subsequences of the inputs with a maximum length of 2000 ms.
The room dimensions and other parameters in \texttt{reverb} are randomly sampled within default ranges~\footnote{https://github.com/DavidDiazGuerra/gpuRIR}.

We evaluate the accuracy of our predictions against target transcriptions using the word error rate (WER). 
The ESB score is the macro-averaged value of datasets in the ESB benchmark, excluding Librispeech.
We implement \wapat on the pre-trained SpeechLM-P~\cite{zhang2022speechlm}, which consists of a Speech Transformer, a Shared Transformer and a CTC head. 
By default, we refer SpeechLM-P-Base to SpeechLM in all tables and figures.
Models are optimized by Adam with a maximum learning rate of 1e-5 and a tri-stage learning rate schedule with the warming-up, holding, and decay periods of [0.1, 0.4, 0.5]. 
We train the models for a total of 30K steps with a batch size of 800 seconds.
Perturbations are bounded with an $l_{\infty}$-norm of $0.01$. 
All experiments are conducted on four NVIDIA Tesla A100.

\subsection{Overall Performance}
To demonstrate the effectiveness of \wapat, we first compared it with data augmentation and adversarial training methods.
We make a fair comparison with standard WavAugment~\cite{park2019specaugment} and SpecAugment~\cite{park2019specaugment}. 
Although SpecAugment performs well on Librispeech test datasets, it shows poor performance in terms of robustness on ESB.
In addition, WavAugment has the suboptimal performance of robustness, with an ESB score of 34.18.
Notably, our \wapat achieves superior performance compared to the above data augmentation methods on both in-domain and out-of-domain datasets by a large margin.
Compared with the waveform space AT method AdvEx~\cite{sun2018training}, \wapat achieves 10.01\% improvement in ESB score.
This further verifies the strengths of our proposed \wapat in terms of generalization on the phoneme space.
Interestingly, \wapat shows obvious advantages on Chime-4 and Common Voice datasets, which share the same speaking style (Narrated) as the LibriSpeech set. 
To provide a more comprehensive evaluation, we test the SpeechLM with the speech enhancement-based method DEMUCS~\cite{defossez2020real}. With sacrificing of some computational efficiency, DEMUCS achieves good performance on generalization, however, still inferior to our method.


\begin{table}[t]  
    \centering
    \caption{Ablation study of the proposed \wapat on cross-domain datasets, (a) is different adversarial training variant, (b) is magnitude $\epsilon$.}
    \vspace{-0.1cm}
    \resizebox{0.99\hsize}{!}{
    \begin{tabular}{lccc}  
    \toprule
    {\multirow{2}*{\textbf{Method}}}& \multicolumn{2}{c}{\textbf{Librispeech}} & {\multirow{2}*{\textbf{ESB Score}}} \\
    ~ & \textbf{test-clean} & \textbf{test-other} & ~ \\
    \midrule
    (a) \textsc{No-AT} & 3.34 & 7.38 & 36.47 \\
     \quad \quad w/ \textsc{Phoneme AT}  & 3.32 & 7.34 & 35.18 \\
     \quad \quad w/ \textsc{WavAugment PAT} & \textbf{3.32} & \textbf{7.28} & \textbf{32.58}\\
    \midrule
    \midrule
    (b) \wapat \\
    \quad \quad $\epsilon$ = \textsc{0.005} & 3.32 & 7.35 & 34.42 \\
    \quad \quad $\epsilon$ = \textsc{0.01}  & \textbf{3.32} & \textbf{7.28} & \textbf{32.58} \\
    \quad \quad $\epsilon$ = \textsc{0.015} & 3.32 & 7.31 & 33.24 \\
    \bottomrule
    \end{tabular}
    }
    \label{tab:ablation}
    \vspace{-0.4cm}
\end{table}

\subsection{Discussion}

We further explore the impact of individual techniques in WavAugment and their combinations with \wapat on the performance of the model, as shown in Figure~\ref{fig:3_2}.
Specifically, we report the percentage of WER reduction for both standard WavAugment and our \wapat, compared to the baseline model.

It can be seen that WavAugment is a useful technique for improving the robustness of models.
However, there are cases where the individual augmentation perform worse than the baseline on certain datasets.
For example, \texttt{time\_mask} increases the WER on the TED-LIUM dataset. 
Furthermore, we note that with the same transformation, the WER reduction of \wapat is greater than that of WavAugment. 
Additionally, for all transformations, there are some oscillations in WavAugment while \wapat is consistently increased compared to the baseline.
The results accords with the expected that phoneme adversarial training with WavAugment guidance constrains stable optimization of adversaries, resulting in better generalization.

\subsection{Ablation Study}
As shown in Table~\ref{tab:ablation}, to better understand the function of each component of \wapat, ablation studies are performed and expected to answer the following questions.

\noindent \textbf{How effective is the PAT?} Echoing (a) in Table~\ref{tab:ablation}, SpeechLM--P with proposed phoneme adversarial training (\textsc{Phoneme AT}) can achieve the better performance on in-domain and out-of-domain datasets than baseline (\textsc{No-AT}). It indicates adversarially altered phoneme perturbations are much closer to the clean distribution, while strengthen the robustness by capturing more robust features.

\noindent \textbf{Is \wapat superior than PAT?}
With \textsc{WavAugment PAT} means that PAT is guided by the WavAugment, \ie, proposed \wapat. The ESB score of \wapat has decreased by roughly 7.4\% when compared to PAT. It is evident that WavAugment guidance AT indeed aids in finding stronger robust features.

\noindent \textbf{Does the choice of magnitude $\epsilon$ matter?}
We present the \wapat results with different magnitude $\epsilon$ in Table~\ref{tab:ablation} (b).
$\epsilon=0$ means the standard training of SpeechLM (\textsc{No-AT}), which makes the models have the worst performance on clean WER and robustness. 
With the increase of $\epsilon$ to 0.01, there is a drop of both clean WER and ESB score.
Moreover, we find the clean WER of target model has the lower sensibility on $\epsilon$.
But with the $\epsilon$ becoming larger, AT greatly damages the generalization, \eg, with $\epsilon=0.015$, ESB score increases to 33.24. This finding is also revealed by~\cite{kireev2022effectiveness}.

\section{Conclusions and Limitations}
In this paper, we propose a novel WavAugment Guided Phoneme Adversarial Training (\wapat) method, to enhance the cross-domain generalization of ASR systems.
\wapat utilizes the phoneme representation of augmented audios to guide the generation of adversarial examples, resulting in consistently stronger generalization on multiple datasets without sacrificing clean performance.
Our experiments demonstrate that \wapat achieves state-of-the-art robustness on challenging ESB benchmark. 
However, \wapat still costs increased training time, this limitation also holds for any adversarial training. This limitation is remained as the future optimization direction.
\bibliographystyle{IEEEtran}
\bibliography{mybib}

\end{document}